\documentclass[showkeys,eqsecnum,showpacs,prd,superscriptaddress,nofootinbib]{revtex4}
\usepackage[dvips]{graphicx}
\usepackage{color}
\usepackage{amsmath}
\usepackage[dvips]{graphicx}
\begin{document}

\title{A Model of Inflation}

\author{Jarmo M\"akel\"a} 

\email[Electronic address: ]{jarmo.makela@puv.fi}  
\affiliation{Vaasa University of Applied Sciences, Wolffintie 30, 65200 Vaasa, Finland}

\begin{abstract}

We consider a novel model of cosmic inflation. In our model one does not need any specific matter field to drive inflation, but inflation stems from the microscopic, Planck scale structure of spacetime, thus being of quantum gravitational origin. At a certain temperature spacetime performs a phase transition, where the cosmological constant drops from a huge, Planck scale value, which is about $10^{87}s^{-2}$ to its present, pretty small value $10^{-35}s^{-2}$. When the cosmological constant is large, the universe goes through a period of very rapid expansion which, however, comes to an abrupt end after the phase transition has been completed. Assuming that the cosmological constant depends of the age of the universe in an appropriate manner during the phase transition one may recover the predictions of the conventional inflationary scenario.  

\end{abstract}


\maketitle

\section{Introduction}

    Many cosmologists believe that very short time after its creation the universe went through a brief period of 
very rapid expansion. This period is known as inflationary epoch, and it is generally thought to have begun, 
when the universe was about $10^{-35}s$ old, and to have lasted about $10^{-33}s$. During the inflationary epoch 
the scale factor of the universe increased at least by a factor $10^{26}$ or so. Originally, cosmic inflation 
was suggested in the 1980's, among others, by Guth \cite{yy} and Linde \cite{kaa} as a solution to certain problems of the 
standard 
cosmological model. \cite{koo} Quite recently, claims of a discovery of direct observational evidence for inflation, based 
on a detailed study of the properties of the cosmic radiation, have been published. \cite{nee} Even though these claims 
have been severely criticized, \cite{vii} and the observational status of inflation is still open, they have brought along 
new interest in the theories of inflation. 

   In most theories of inflation one assumes that the very early universe was filled with a scalar field, which 
is known as inflaton field. With an appropriate choice of the parameters of the inflaton field one finds that 
Einstein's fleld equation possesses solutions, which are qualitatively similar to the one describing the 
so-called de Sitter universe, which is an empty universe with  constant spatial curvature and positive cosmological constant, and where the 
scale factor increases exponentially as a function of the age of the universe. During the inflationary epoch the 
inflaton field is effectively constant producing an exponential increase in the scale factor, but after a while 
the inflaton field rapidly decreases, and the inflation comes to an end. \cite{koo}

    In this paper we shall consider a novel model of inflation. In our model there is no need to introduce specific matter fields to drive inflation, but the inflationary expansion of the universe stems from the microscopic, Planck scale structure of spacetime. In this sense inflation is thought to be of quantum gravitational origin. Our model is based on the results of Ref. \cite{kuu}. In Ref. \cite{kuu} it was found that if one constructs the so-called shrinked horizon of the de Sitter spacetime - which is a certain spacelike two-sphere just inside of the cosmological horizon of the de Sitter spacetime - out of discrete constituents, each of these constituents contributing to the shrinked horizon an area, which is an integer times a constant, an explicit, analytic expression may be found for the partition function of the 
de Sitter spacetime. The partition function implies, among other things, that at a certain temperature the de Sitter spacetime performs a {\it phase transition},  where the cosmological constant $\Lambda$ drops from a huge Planck scale value, which is around $10^{87}s^{-2}$, to its presently observed value, which is around $10^{-35}s^{-2}$. Our idea is to explain inflation by means of this property of the de Sitter spacetime: Immediately after the creation of the universe the cosmologal constant $\Lambda$ was about $10^{87}s^{-2}$, and a short time afterwards a period of extremely rapid expansion took place. This period came to an end, when the universe performed a phase transition from the high to the low $\Lambda$ phase. The cosmological constant dropped by about 122 orders of magnitude, and matter and radiation were created out of the huge vacuum energy of the universe.

   Of crucial importance is the question of how fast did the cosmological constant decrease during the phase transition. In this paper we consider a model, where  the cosmological constant is inversely proportional to the square of the age of the universe. Such model is attractive on dimensional grounds, because the SI unit of the cosmological constant is $s^{-2}$. It turns out that the constants of integration in the resulting solution to Einstein's field equation equipped with a varying cosmological constant may be chosen in such a way that the predictions of the conventional inflationary scenario are recovered. The possiblity that inflation might be caused by quantum gravitational effects  and the discreteness of spacetime has also been considered by Bojowald {\it et al.} \cite{kuu1} However, that approach is based on loop quantum gravity, and it is completely different from the approach employed in this paper.  The idea of a connection between the expansion of the universe and the possible discreteness of spacetime has been investigated by Padmanabhan. \cite{kuu2}

   Unless otherwise stated, we shall always use the natural units, where $\hbar = c = G = k_B = 1$.

  \section{Preliminaries}

   In a model, where the cosmic inflation of the universe is driven by a varying cosmological constant the universe may be described, effectively, by the de Sitter spacetime {\footnote{There is a mixture of terminologies here: In most textbooks the de Sitter universe is defined to be flat, whereas some authors consider the de Sitter universe as closed, which is actually the closed, empty Lemaitre universe. Unless otherwise stated we shall, in this paper, always assume implicitly that the de Sitter universe is closed. However, the results are essentially unchanged, no matter, whether the universe is flat or closed. In particular, the results reviewed in this section do not depend on, whether the de Sitter universe is flat or closed.}} during the early stages of its evolution. In the static coordinates the line element of the de Sitter spacetime may be written as: \cite{seite}
\begin{equation}
ds^2 =- \left(1 - \frac{\Lambda}{3}r^2\right)\,dt^2 + \frac{dr^2}{1 - \frac{\Lambda}{3}r^2} + r^2\,d\theta^2 + r^2\sin^2\theta\,d\phi^2,
\end{equation}
where $\Lambda>0$ is the cosmological constant, and $r$, $\theta$ and $\phi$ are the spherical coordinates. The de Sitter spacetime has the {\it cosmological horizon}, where
\begin{equation}
r = r_C := \sqrt{\frac{3}{\Lambda}}.
\end{equation}
The only non-vanishing component of the future pointing unit tangent vector field $u^\mu$ of the congruence of the world lines of observers with constant $r$, $\theta$ and $\phi$ in the de Sitter spacetime, when $r < r_C$, is:
\begin{equation}
u^t = \left(1 - \frac{\Lambda}{3}r^2\right)^{-1/2},
\end{equation}
and one finds that the norm of the proper acceleration vector field 
\begin{equation}
a^\mu := u^\alpha u^\mu_{;\alpha}
\end{equation}
of this congruence is:
\begin{equation}
a := \sqrt{a_\mu a^\mu} = \left(1 - \frac{\Lambda}{3}r^2\right)^{-1/2}\frac{\Lambda}{3}r.
\end{equation}

    Eq. (2.5) gives the proper acceleration of an observer with constant static coordinates $r$, $\theta$ and $\phi$ in the de Sitter spacetime such that $r < r_C$. In Ref. \cite{kuu} the cosmological "constant" $\Lambda$ was not assumed to be a constant, but a free thermodynamical variable, and the thermodynamics of the de Sitter spacetime was considered from the point of view of an observer with constant proper acceleration $a$ just inside of the cosmological horizon, where $r = r_C$. For the sake of brevity and simplicity the spacelike two-sphere with constant $r < r_C$ in the de Sitter spacetime was called as a {\it shrinked horizon}, and it was assumed that when the cosmological constant $\Lambda$ varies, the radius $r$ of the shrinked horizon will also vary, but in such a way that the proper acceleration $a$ of an observer at rest on the shrinked horizon is kept as a constant. Among other things, it was found that the quantity
\begin{equation}
E := \frac{a}{8\pi}A,
\end{equation}
where $A := 4\pi r^2$ is the area of the shrinked horizon, describes the total energy of the de Sitter spacetime from the point of view of such observer.

    The thermodynamical properties of any system arise from its microscopic properties. In Ref. \cite{kuu} the shrinked horizon of the de Sitter spacetime was constructed out of discrete constituents, each of them contributing to the shrinked horizon an area, which is an integer times a constant. As a consequence, the area of the shrinked horizon takes the form:
\begin{equation}
A = \alpha\ell^2_{Pl}(n_1 + n_2 +... + n_N),
\end{equation}
where $n_1, n_2,..., n_N$ are non-negative integers, $\alpha$ is a pure number to be determined later, and 
\begin{equation}
\ell_{Pl} := \sqrt{\frac{\hbar G}{c^3}} \approx 1.6\times 10^{-35}m
\end{equation}
is the Planck length. In Eq. (2.7) $N$, which is assumed to be very large and fixed, gives the  number of the constituents of the shrinked horizon. The non-negative integers $n_j$ $(j = 1, 2,..., N)$ are quantum numbers determining the quantum states of the constituents of the shrinked horizon. Constituent $j$ is in vacuum, if $n_j = 0$; otherwise the constituent is in an excited state. Eqs. (2.6) and (2.7) imply that the possible energies of the de Sitter spacetime from the point of view of an observer on the shrinked horizon, where $a = constant$ are of the form:
\begin{equation}
E_n = n\alpha\frac{a}{8\pi},
\end{equation}
where
\begin{equation}
n : = n_1 + n_2 + .... + n_N.
\end{equation}
Taking the number $g(E_n)$ of the degenerate states associated with the same energy $E_n$ to be the number of the different combinations of the non-vacuum quantum states of the constituents of the shrinked horizon yielding the same energy one may obtain an explicit, analytic expression for the partition function 
\begin{equation}
Z(\beta) := \sum_n g(E_n)e^{-\beta E_n}
\end{equation}
of the de Sitter spacetime from the point of view of our observer. Defining the {\it characteristic temperature}
\begin{equation}
T_C := \frac{\alpha a}{8\pi\ln 2}
\end{equation}
the result turns out to be: \cite{kuu}
\begin{equation}
Z(\beta) = \frac{1}{2^{\beta T_C}- 2}\left[1 - \left(\frac{1}{2^{\beta T_C} - 1}\right)^{N+1}\right],
\end{equation}
when $\beta T_C \ne 1$, and 
\begin{equation}
Z(\beta) = N+1,
\end{equation}
when $\beta T_C = 1$.  

    The thermodynamical properties of the de Sitter spacetime may be deduced from Eqs. (2.13) and (2.14). The most interesting of them concerns the quantity
\begin{equation}
\bar{E}(\beta) := - \frac{1}{N}\frac{\partial}{\partial\beta}\ln Z(\beta),
\end{equation}
which is the average energy per constituent of the de Sitter spacetime. Using Eq. (2.13) one finds that in the leading approximation for large $N$:
\begin{equation}
\bar{E}(\beta) = \left[\frac{1}{N}\frac{2^{\beta T_C}}{2^{\beta T_C} - 2} + \frac{2^{\beta T_C}}{2^{\beta T_C} - 1 - (2^{\beta T_C} - 1)^{N+2}}\right]T_C\ln 2.
\end{equation}
As one may observe, something very curious happens, when 
\begin{equation}
T := \frac{1}{\beta} = T_C.
\end{equation}
One notes that 
\begin{equation}
\lim_{N\rightarrow\infty}\bar{E}(\beta) = 0,
\end{equation}
when $T < T_C$, and 
\begin{equation}
\lim_{N\rightarrow\infty}\bar{E}(\beta) = \frac{2^{\beta T_C}}{2^{\beta T_C} - 1}T_C\ln 2,
\end{equation}
when $T > T_C$. Eqs. (2.18) and (2.19) imply that de Sitter spacetime performs, according to our discrete model of the shrinked horizon, a {\it phase transition} at the characteristic temperature $T_C$: According to Eq. (2.18) the constituents of the shrinked horizon are effectively in vacuum, when $T < T_C$, whereas using Eqs. (2.9), (2.10), (2.12) and (2.19) we find that the constituents have jumped, in average, to the second excited states, where $n_j = 2$, when $T$ is slightly higher than $T_C$. The latent heat per constituent associated with this phase transition is
\begin{equation}
\bar{L} = 2T_C\ln 2.
\end{equation}
It is interesting to note that the characteristic temperature $T_C$ in Eq. (2.12) agrees with the Unruh temperature $T_U := \frac{a}{2\pi}$ of the observer on the shrinked horizon, where the proper acceleration $a = constant$, if we put
\begin{equation}
\alpha = 4\ln 2.
\end{equation}
With this choice for the constant $\alpha$ one finds, among other things, that the entropy of the de Sitter spacetime takes, in the natural units, the form: \cite{kuu}
\begin{equation}
S = \frac {1}{4}A,
\end{equation}
when $T = T_C$. Since the cosmological horizon is assumed to lie just inside of the cosmological horizon of the de Sitter spacetime we may identify, for all practical purposes, the shrinked horizon area $A$ with the area of the cosmological horizon. Hence our discrete model reproduces the classic result of Gibbons and Hawking. \cite{kasi} It should be noted that models similar to the one constructed here for the shrinked horizon of the de Sitter spacetime were constructed for the stretched horizon of the Schwarzschild black hole in Ref. \cite{ysi} and for that of the Reissner-Nordstr\"om hole in Ref. \cite{kymppi}. In both cases the discrete model  implies, among other things, that the entropy of a black hole must be one-quarter of its event horizon area.

\section{The Model}

        The results reviewed in Section 2 have profound implications on the properties of the cosmological constant and the evolution of the universe. Employing Eqs. (2.2) and (2.6) we find that the cosmological constant $\Lambda$ may be expressed in terms of the energy $E$ of the de Sitter spacetime as:
\begin{equation}
\Lambda = \frac{3a}{2E},
\end{equation}
and if we use Eqs. (2.12), (2.16) and (2.21), we may write $\Lambda$ as a function of the temperature parameter $\beta = \frac{1}{T}$ and the number $N$ of the constituents of the shrinked horizon, in the SI units, as:
\begin{equation}
\Lambda = \left[\frac{2^{\beta E_C}}{2^{\beta E_C} - 2} + N\frac{2^{\beta E_C}}{2^{\beta E_C} - 1 - (2^{\beta E_C} - 1)^{N+2}}\right]^{-1}\frac{3\pi}{\ln 2}\frac{c^2}{\ell_{Pl}^2},
\end{equation}
where we have denoted:
\begin{equation}
E_C :=k_B T_C.
\end{equation}
In the limit, where $T\rightarrow 0$, the temperature parameter $\beta\rightarrow\infty$, and we find:
\begin{equation}
\lim_{T\rightarrow 0}\Lambda = \frac{3\pi}{\ln 2}\frac{c^2}{\ell_{Pl}^2}\approx 4.8\times 10^{87}s^{-2},
\end{equation}
which is very large, indeed. Actually, the right hand side of Eq. (3.4) is the numerical value one expects for the cosmological constant, if one interprets the cosmological constant $\Lambda$ as a quantity proportional to the vacuum energy density of the matter fields. As far as $T < T_C$, the expression given by Eq. (3.2) for $\Lambda$ is essentially a constant function of $T$. At the characteristic temperature $T_C$, however, the cosmological constant will drop drastically. When $T > T_C$, the cosmological constant is given, in effect, by an expression:
\begin{equation}
\Lambda = \frac{2^{\beta E_C} - 1}{2^{\beta E_C}}\frac{1}{N}\frac{3\pi}{\ln 2}\frac{c^2}{\ell_{Pl}^2},
\end{equation}
and one finds that
\begin{equation}
\lim_{T\rightarrow T_C^+}\Lambda = \frac{3\pi}{2N\ln 2}\frac{c^2}{\ell_{Pl}^2},
\end{equation}
which is the cosmological constant immediately after the phase transition in the de Sitter spacetime has been completed. Taking
\begin{equation}
N \sim 10^{122}
\end{equation}
we observe that
\begin{equation}
\Lambda \sim 10^{-35}s^{-2},
\end {equation}
which is the current estimate, based on observations, for the cosmological constant. To summarize, we have therefore shown that when the temperature of the universe from the point of view of an observer on the shrinked horizon $a = constant$ exceeds the characteristic temperature $T_C$, the cosmological constant drops, according to our model, from its huge, Planck-scale value to its presently observed, pretty small value.

    The expansion of the de Sitter spacetime is essentially exponential with an e-folding time $\sqrt{\frac{3}{\Lambda}}$ in the comoving time $\tau$, and hence large cosmological constant $\Lambda$ implies very rapid expansion. In very short terms we may therefore attempt to explain the cosmic inflation of the very early universe as follows: Immediately after its creation the universe was in a state of very low entropy and temperature, and the cosmological constant was given by the right hand side of Eq. (3.4). As a consequence, the universe went through a period of  very rapid expansion which, however, came to an abrupt end, when the temperature of the universe got sufficiently large for a phase transition to take place in the universe, and the cosmological constant dropped to its present, pretty small value. An attractive feature of the model outlined above is that it provides for the cosmic  inflation a very simple and natural explanation without any need to introduce specific matter fields to drive inflation. In our model the underlying reason for inflation may be traced back to the fundamentally discrete structure of spacetime. However, there are several delicate details in the model, which are of crucial importance for the behaviour of the universe during the inflationary epoch.

     The key question is: In which way did the transition from the high to the  $\Lambda$ phase take place? Did it take place immediately after the inflationary epoch, or already - at least in parts -  during the inflation? If the phase transition happened already during the inflation, then how fast did the cosmological constant $\Lambda$ drop during the inflationary expansion of the universe?

    Let us first consider the very extreme possibility, where the phase transition did not take place during the inflation, but only afterwards, by means of an extremely rapid process, thus putting an end to the inflation. In that case the cosmological constant $\Lambda$ really is a constant during the inflation, and its numerical value is given by the right hand side of Eq. (3.4). The scale factor $R$ of the universe obeys, as a function of the comoving time $\tau$, an equation:
\begin{equation}
R(\tau) = R_0 \cosh(\sqrt{\frac{\Lambda}{3}}\tau),
\end{equation}
where
\begin{equation}
R_0 := \sqrt{\frac{3c^2}{\Lambda}}.
\end{equation}
Eq. (3.9) implies that the time needed for the scale factor of the universe to decrease from $R_0$ to $R_1$ is:
\begin{equation}
\tau = \sqrt{\frac{3}{\Lambda}}\cosh^{-1}(\frac{R_1}{R_0}).
\end{equation}
Assuming that 
\begin{equation}
R_1 \sim 10^{26}m,
\end{equation}
which is the radius of the presently observable universe we find, putting for $\Lambda$ the value given by the right hand side of Eq. (3.4):
\begin{equation}
\tau \sim 10^{-42}s.
\end{equation}
Hence our model implies, in this very extreme case, that the present universe was created out of nothing in about Planck time. Once after the universe reached its present size, the cosmological constant dropped to its present value, and matter, radiation and everything we see around us were suddenly created out of the vacuum energy of the universe. Such possibility, where the universe was created more or less as we observe it today in an almost zero time is untenable for almost  any practising cosmologist.

      We must therefore turn our attention to another possibility, which is that the cosmological constant was not a constant during the inflation, but already during the inflation it began a decrease towards its present value. Cosmological solutions to Einstein's field equation in the presence of a varying cosmological constant have been studied, among others, by Overduin and Cooperstock.   
   \cite{yytoo} Their idea was to write Einstein's field equation with a cosmological constant as:
\begin{equation}
G_ {\mu\nu} = 8\pi\tilde{T}_{\mu\nu},
\end{equation}
where
\begin{equation}
\tilde{T}_{\mu\nu} := T_{\mu\nu} - \frac{1}{8\pi}\Lambda g_{\mu\nu}.
\end{equation}
When Einstein's field equation is written in this way, the cosmological constant $\Lambda$ may be thought as a part of the matter content of the universe, rather than as a purely geometrical quantity: When the cosmological constant $\Lambda$ decreases, the energy associated with $\Lambda$ and thus the vacuum energy of the universe is converted to the energy of matter and radiation.  Using Einstein's field equation Overduin and Cooperstock managed to obtain for the scale factor $R(\tau)$ of the universe in the presence of a varying cosmological constant an equation:
\begin{equation}
\frac{\ddot{R}}{R} = \left(1 - \frac{3\gamma}{2}\right)\left(\frac{\dot{R}^2}{R^2} + \frac{k}{R^2}\right) + \frac{\gamma}{2}\Lambda.
\end{equation}
In this equation $k=1$, $0$ or $-1$, depending on whether the universe is closed, flat or open, respectively, and the pure number $\gamma$ relates the pressure $p$ of the universe to its energy density $\rho$ as:
\begin{equation}
p = (\gamma - 1)\rho.
\end{equation}
In the matter-dominated universe, for instance, $\gamma = 1$, whereas in the radiation-dominated universe $\gamma = 4/3$. The dot means the derivative with respect to the comoving time $\tau$. 

    We may assume that during its very early stages of evolution the universe was  radiation-dominated. With this assumption we may put $\gamma = 4/3$, and Eq. (3.16) reduces, in the SI units, to the form:
\begin{equation}
\ddot{y} - \frac{4}{3}\Lambda y = -2kc^2,
\end{equation}
where we have denoted:
\begin{equation}
y := R^2.
\end{equation}
As an example of a cosmological model, where the cosmological constant $\Lambda$ begins to decrease towards its present value already during the inflationary epoch, let us assume that during the very early stages of the evolution of the universe the cosmological constant decreased as a function of the comoving time $\tau$ as:
\begin{equation}
\Lambda(\tau) = \frac{6n^2-3n}{2\tau^2},
\end{equation}
where the pure number $n > 1/2$ is a constant. One readily finds that with these assumptions Eq. (3.18) has the solution:
\begin{equation}
y(\tau) = \tilde{R}^2(\frac{\tau}{\tau_0})^{2n} +  k\frac{(c\tau)^2}{2n^2 - n - 1},
\end{equation}
where $\tilde{R}$ and $\tau_0$ are positive constants. Hence Eq. (3.19) implies the following time dependence for the scale factor of the universe:
\begin{equation}
R(\tau) = \sqrt{\tilde{R}^2(\frac{\tau}{\tau_0})^{2n} +  k\frac{(c\tau)^2}{2n^2 - n - 1}}.
\end{equation}

     Consider first the case, where the universer is flat. In that case $k = 0$, and Eq. (3.22) reduces to the form:
\begin{equation}
R(\tau) = \tilde{R}(\frac{\tau}{\tau_0})^n.
\end{equation}
According to the conventional inflationary scenario the inflation began, when the universe was about $10^{-35}s$ old, and it lasted about $10^{-33}s$. During the inflation the scale factor of the universe is supposed to have increased at least by a factor $10^{26}$ or so, even though much larger factors have been suggested. \cite{koo} With these assumptions we find that Eq. (3.23) implies:
\begin{equation}
\left(\frac{10^{-33}}{10^{-35}}\right)^n \sim 10^{26},
\end{equation}
and so we must have:
\begin{equation}
n \approx 13.
\end{equation}
When $n$ is this large, the function $R(\tau)$ in Eq. (3.23) increases pretty slowly until $\tau \sim \tau_0$, but when $\tau$ gets above $\tau_0$, $R(\tau)$ begins to icrease very rapidly. If the inflation began, when $\tau \sim 10^{-35}s$, we must therefore have:
\begin{equation}
\tau_0 \sim 10^{-35}s.
\end{equation}

      It is interesting that our expression in Eq. (3.20) for the cosmological constant $\Lambda(\tau)$ depends on the power $n$ only, being independent of both the constant $\tau_0$, which gives the age of the universe, when the inflation began, and of the constant $\tilde{R}$, which determines the size of the universe at the onset of inflation. Putting $n = 13$ in Eq. (3.20) we find:
\begin{equation}
\Lambda(\tau) = \frac{975}{2\tau^2}.
\end{equation}
We may assume that Eq. (3.27) gives a phenomenological description of the time evolution of the cosmological constant from the very early stages of the evolution of the universe to the end of inflation. It is commonly believed that it makes no sense to talk about times earlier than the Planck time, which is about $10^{-43}s$, after the creation of the universe. Putting $\tau \sim 10^{-43}s$ in Eq. (3.27) we get:
\begin{equation}
\Lambda \sim 10^{88}
\end{equation}
which is consistent with Eq. (3.4). Hence Eq. (3.27) implies that the cosmological constant $\Lambda$ began with its Planck scale value  given by the right hand side of Eq. (3.4) at the creation of the universe, and immediately after its creation the universe began a phase transition from the high to the low $\Lambda$ phase. Putting $\tau \sim 10^{-33}s$ in Eq. (3.27) we find that $\Lambda \sim 10^{68}s^{-2}$, and so we may conclude that even though the cosmological constant had already gone down by about 20 orders of magnitude at the end of inflation, the phase transition was still far from being completed. To make our model consistent with the conventional inflationary scenario, where the inflation came to an end, when the unverse was about $10^{-33}s$ old, we must therefore assume that for a still unknown reason the rate of the phase transition  increased drastically, when the age of the universe exceeded $10^{-33}s$.

    When the universe is closed, $k = 1$, and the results found above remain otherwise unchanged, except that when $\tau < \tau_0$, the scale factor depends, in effect, linearly on the age of the universe. However, if the universe is open, $k = -1$, and the solution  given by Eq. (3.22) is valid only if
\begin{equation}
\tau \ge \tilde{\tau} := \left(\frac{c\tau_0}{\tilde{R}}\frac{1}{\sqrt{2n^2 - n - 1}}\right)^{\frac{1}{n-1}}\tau_0,
\end{equation}
because otherwise $R(\tau)$ would become imaginary. For large $n$ the prefactor in front of $\tau_0$ is of the order of unity, whenever $\tilde{R}$ is within physically reasonable bounds, and therefore $\tilde{\tau} \sim \tau_0$. When $\frac{\tau}{\tau_0} \gg 1$, Eq, (3.22) implies that
\begin{equation}
R(\tau) \sim \tilde{R}(\frac{\tau}{\tau_0})^n,
\end{equation}
and so we recover, in effect, the results obtained for the flat universe during the inflation.

\section{Concluding Remarks}

  In this paper we have considered a model of inflation, where the inflationary expansion of the very early universe stems from the properties of the cosmological constant alone, without any need to introduce specific matter fields to drive inflation. In our model the underlying origin of inflation lies in the fundamentally discrete structure of spacetime: Assuming that the so-called shrinked horizon of the de Sitter spacetime, which is a certain spacelike two-sphere just inside of the cosmological horizon of the de Sitter spacetime, consists of discrete constituents, each of them contributing to the shrinked horizon an area, which is an integer times a constant, one finds that at a certain temperature the de Sitter spacetime performs a phase transition, where the cosmological constant $\Lambda$ drops from a Planck scale value $10^{87}s^{-2}$ to its presently observed value, which is around $10^{-35}s^{-2}$. During the era, when the cosmological constant was very large, the universe went through a period of very rapid expansion which, however, came to an abrupt end once after the phase transition from the high to the low $\Lambda$ phase was completed. Originally, the universe was an empty de Sitter universe, but when the cosmological constant decreased, the energy needed for the creation of the matter and the radiation of the universe was released from the vacuum energy of the universe. 

   To reproduce the predictions of the conventional inflationary scenario one has to make assumptions on the rate of the phase transition from the high to the low $\Lambda$ phase. In this paper we considered a model, where the phase transition began immediately after the creation of the universe, and the cosmological constant was, until the end of inflation, inversely proportional to the square of the age of the universe. With these assumptions Einstein's field equation equipped with a varying cosmological constant has a solution, where the scale factor of the universe obeys, during the inflation, a simple power law as a function of its age. The predictions of the conventional inlationary scenario are reproduced, if the scale factor is proportional to the thirteenh power of the age of the universe, and this power is consistent with an appropriate choice of the constant of proportionality in the expression of $\Lambda$.

   Even though our model provides a potential explanation for inflation, it still has many problems, and several questions are left open. In particular, we do not know exactly how fast the phase transition in spacetime took place. Even though a model, where the cosmological constant is inversely proportional to the square of the age of the universe is attractive on dimensional grounds (the SI unit of the cosmological constant is $s^{-2}$), there is not any particular reason for such dependence of the cosmological constant on time. Moreover, the parameters of the associated solution to Einstein's field equation were chosen on purely {\it ad hoc} grounds, keeping in mind that the solution should describe inflation. The solution of these problems, however, would require a detailed knowledge on the quantum-mechanical interaction of radiation and spacetime in the early universe. Such knowledge will be missing until a satisfactory quantum theroy of gravitation will be found. Nevertheless, it would be very interesting, if cosmic inflation really stemmed from a microscopic structure of spacetime, thus being of quantum gravitational origin. In that case attempts to observe effects predicted by inflation in the present universe would actually be attempts to observe quantum effects of gravitation.


\begin{thebibliography} {20}

\bibitem{yy} A. H. Guth, Phys. Rev. {\bf  D23}, 347 (1981).

\medskip

\bibitem{kaa} A. Linde, Phys. Lett. {\bf  B108}, 389 (1982).

\medskip

\bibitem{koo} For a sample of general reviews of inflation see, for example, S. Tsujikawa, hep-ph/0304257, S. Watson, astro-ph/0005003, D. Baumann, arXiv:0907.5424, A. Linde, Lect. Notes Phys. {\bf 738}, 1 (2008) (arXiv:0705.0164). 

\medskip

\bibitem{nee} P. A. R. Ade {\it et al.}. Phys. Rev. Lett. {\bf 112}, 241101 (2014).

\medskip

\bibitem{vii} See, for example, R. Flauger, J. C. Hill and D. N. Spergel, arXiv:1405.7351.

\medskip

\bibitem{kuu} J. M\"akel\"a, Phys. Rev. {\bf D87}, 104040 (2013).

\medskip

\bibitem{kuu1} M. Bojowald, Phys. Rev. Lett. {\bf 39}, 261301 (2002), M. Bojowald and K. Vandersloot, Phys. Rev. {\bf D67}, 124023 (2003), M. Bojowald, G. Calcagni and S. Tsujikawa, JCAP {\bf 111}, 046 (2011).

\medskip

\bibitem{kuu2} T. Padmanabhan, arXiv:1206.4956, arXiv:1208.1375, arXiv:1210.4174.

\medskip

\bibitem{seite} See, for example, R. Bousso, hep-th/0205177.

\medskip

\bibitem{kasi} G. W. Gibbons and S. W. Hawking, Phys. Rev. {\bf D15}, 2738 (1977).

\medskip

\bibitem{ysi} J. M\"akel\"a. Entropy {\bf 13}, 1324 (2011).

\medskip

\bibitem{kymppi} J. M\"akel\"a, Int. J. Mod. Phys. {\bf D23}, 145001 (2014).

\medskip

\bibitem{yytoo} J. M. Overduin and F. I. Cooperstock, Phys. Rev. {\bf D58}, 043506 (1998).

\end{thebibliography}
\end{document}